\documentclass[aps,prd,twocolumn,fleqn,floatfix]{revtex4}

\usepackage{graphicx}
\usepackage{amsmath,amssymb,amsfonts}
\usepackage{color}

\newcommand{\mD}{m_\mathrm{D}}
\newcommand{\qT}{\hat{q}/T^3}
\newcommand{\qmax}{q_\mathrm{max}}

\begin{document}

\title{$\eta/s - \qT$ Relation at Next-to-Leading Order in QCD}

\author{Berndt M\"uller}
\affiliation{Department of Physics, Duke University, Durham, NC 27708}

\date{\today}

\begin{abstract}
The relation between the specific shear viscosity $\eta/s$ and the dimensionless jet quenching parameter $\qT$ in perturbative QCD is explored at next-to-leading order in the coupling constant. It is shown that the relation changes little, although both transport coefficients independently are subject to large modifications at the NLO level. This finding confirms that the relationship is robust.
\end{abstract}

\maketitle

The shear viscosity $\eta$ and the so-called jet quenching parameter $\hat{q}$ are two important transport coefficients that characterize the quark-gluon plasma. In order to eliminate the trivial temperature dependence of these quantities, it is customary to consider the two dimensionless quantities $\eta/s$, the ratio of the shear viscosity and the entropy density, sometimes called the specific shear viscosity, and $\qT$, where $T$ is the temperature. Both dimensionless quantities are sensitive to the effective coupling strength in the plasma. When the coupling is weak, $\eta/s$ is large and $\qT$ is small; when the coupling is strong, $\eta/s$ is small and $\qT$ is large. The fact that the analysis of experimental data from relativistic heavy ion collisions indicates that $\eta/s \approx 0.1-0.2$ \cite{Bernhard:2019bmu,Everett:2020xug} and $\qT \approx 3-4$ \cite{Burke:2013yra,Cao:2021keo} is generally taken as an indication that the quark-gluon plasma produced in such collisions is strongly coupled.

Some time ago, Majumder, M\"uller, and Wang (MMW) argued that the two quantities are connected by a general relation of the form \cite{Majumder:2007zh}:
\begin{equation}
\frac{\eta}{s} \cdot \frac{\hat{q}}{T^3} \approx \mathrm{const.} 
\label{eq:MMW}
\end{equation}
in any perturbative gauge theory where small-angle scattering is dominated by the exchange of massless or low-mass quanta. Using lowest-order QCD perturbation theory they estimated the constant $C$ to be of order unity and independent of the gauge coupling $g$.

Since we are now in possession of experimentally derived estimates for the two quantities, it makes sense to revisit the MMW relation and explore to what extent it remains valid at higher orders of perturbation theory. This exploration is aided by recent calculations of the transport coefficients $\eta$ and $\hat{q}$ at next-to-leading order (NLO) of thermal perturbation theory \cite{Ghiglieri:2015ala,Ghiglieri:2018dib}. 

Before presenting the details, it makes sense to revisit the argument why the relation (\ref{eq:MMW}) should have rather general validity. The argument assumes that the medium can be described by quasiparticles and that the total scattering cross section between quasiparticles is dominated by small-angle scatterings as is generally the case when the interactions among quasiparticles are mediated by massless or low-mass quanta, such as in QCD. 

The transport coefficient $\hat{q}$ governing the radiative energy loss of a propagating parton in SU(3)-color representation $R$ is given by \cite{Baier:1996kr}:
\begin{equation}
\hat{q}_R = \tilde\rho \int d^2q_\perp\, q_\perp^2 \frac{d\sigma_R}{d^2q_\perp} \, ,
\label{eq:qhat-sigma}
\end{equation}
where  $d\sigma/d^2q_\perp$ denotes the differential cross section of elastic scattering for an energetic parton on medium constituents and
\begin{equation}
\tilde\rho = \nu \int \frac{d^3k}{(2\pi)^3} n(k) [1\pm n(k)]
\end{equation}
is the final-state weighted density of quasiparticles in the medium. $\nu$ denotes the quasiparticle degeneracy and the sign distinguishes between bosons and fermions. The lowest-order, leading-logarithmic (LL) result for a quark-gluon plasma is \cite{Ghiglieri:2015ala}:
\begin{equation}
\hat{q}_R^\mathrm{(LL)}(\qmax) = \frac{g^2 C_R T \mD^2}{4\pi} \ln\frac{\qmax^2}{\mD^2} \, ,
\label{eq:qhatLL}
\end{equation}
where $C_R$ is the quadratic Casimir for the color representation of the fast parton that initiates the jet, $\mD$ is the Debye screening mass in the plasma; and $\qmax$ is an ultraviolet cut-off (see \cite{CaronHuot:2008ni}). Here we choose $\qmax = \mu$ with $\mu/T = 2.765$ for the pure gauge theory and $\mu/T = 2.957$ for three-flavor QCD. This choice makes the logarithm consistent with the one encountered in the calculation of $\eta/s$. Other possible process-dependent choices are discussed in \cite{CaronHuot:2008ni}.

The shear viscosity $\eta$ is defined as the coefficient of the contribution to the stress tensor of the medium that is proportional to the divergence-free part of the velocity gradient. In the framework of kinetic theory based on a quasi-particle picture, the shear viscosity is given in terms of the mean-free path $\lambda_{\rm f}(p)$ of a constituent particle of momentum $p$:
\begin{equation}
\eta \approx \kappa \rho \langle p \rangle \lambda_{\rm f} ,
\label{eq:visc}
\end{equation}
where $\rho$ is the density of medium constituents (gluons and quarks) and $\kappa \approx 1/3$ \cite{Danielewicz:1984ww}. 

The relation between $\eta$ and $\hat{q}$ is derived from the observation that the mean-free path is related to the average transport cross section of a quasi-particle in the medium:  $\lambda_\mathrm{f} = (\tilde\rho\,\sigma_\mathrm{tr})^{-1}$. When small-angle scattering is dominant, the transport cross section is related to the differential elastic scattering  cross section by the relation \cite{Danielewicz:1984ww}:
\begin{equation}
\sigma_\mathrm{tr} = \int d\Omega \frac{d\sigma}{d\Omega} \sin^2\theta
\approx \frac{4}{s} \int d^2q_{\perp}\, q_{\perp}^2\, \frac{d\sigma}{d^2q_{\perp}} \, .
\label{eq:sigma-tr}
\end{equation}
where $\sqrt{s}$ is the center-of-mass energy of the quasiparticle collision. Comparing with (\ref{eq:qhat-sigma}) we find:
\begin{equation}
\lambda_\mathrm{f} \approx \frac{s}{4 \hat{q}}  \, .
\label{eq:sigma-tr-qhat}
\end{equation}
For a thermal ensemble of light particles (gluons and light quarks), $\langle p \rangle \approx 3T$ and $\langle s \rangle \approx 18 T^2$, and thus:
\begin{equation}
\eta \approx 13.5\, \kappa \frac{\rho T^3}{\hat{q}} .
\label{eq:eta-qhat-approx}
\end{equation}
Applying the expression $s \approx 3.6\,\rho$ for the entropy density of a gas of free gluons, a relation of the MMW type is obtained (see Eq.~(5) in \cite{Majumder:2007zh}):
\begin{equation}
\frac{\eta}{s} \approx 3.75 \kappa \frac{T^3}{\hat{q}_A} \approx 1.25 \frac{T^3}{\hat{q}_A} \, .
\label{eq:eta-s-QP}
\end{equation}

The exact result for QCD at LL order in the QCD coupling constant for a pure gluon gas is \cite{Arnold:2000dr}:
\begin{equation}
\eta_\mathrm{g} = \frac{0.343~T^3}{\alpha_s^2 \ln(1/\alpha_s)} \, .
\label{eq:etag-LL}
\end{equation}
With the entropy density of a gas of free gluons given by $s_\mathrm{g} = (32\pi^2/45) T^3$ the specific shear viscosity is:
\begin{equation}
\frac{\eta_\mathrm{g}}{s_\mathrm{g}} = \frac{0.0489}{\alpha_s^2 \ln(1/\alpha_s)} \, .
\label{eq:etag-s-LO}
\end{equation}
Inserting the Debye mass for a gluon gas, $\mD^2 = 4\pi\alpha_s T^2$, into the lowest-order result for the jet quenching parameter (\ref{eq:qhatLL}) for a gluon, one finds:
\begin{equation}
\frac{\hat{q}_A}{T^3} = 12 \pi \alpha_s^2 \ln(1/\alpha_s) \, ,
\label{eq:qhatT3-g}
\end{equation}
giving the exact LL order MMW relation for a pure gluon gas:
\begin{equation}
\frac{\eta_\mathrm{g}}{s_\mathrm{g}} \cdot \frac{\hat{q}_A}{T^3} = 1.84 \, .
\label{eq:MMWg}
\end{equation}

Repeating the calculation for a three-flavor ($N_f = 3$) quark-gluon plasma and a quark jet, one obtains:
\begin{equation}
\left(\frac{\eta}{s}\right)_{N_f=3} = \frac{0.06474}{\alpha_s^2 \ln(1/\alpha_s)} \, ,
\label{eq:etaQGP-s-LO}
\end{equation}
\begin{equation}
\frac{\hat{q}_F}{T^3} = 8 \pi \alpha_s^2 \ln(1/\alpha_s) \, ,
\label{eq:qhatT3-QGP}
\end{equation}
Combining the two results, one finds:
\begin{equation}
\left(\frac{\eta}{s}\right)_{N_f=3} \cdot \frac{\hat{q}_F}{T^3} = 1.63 \, .
\label{eq:MMW3}
\end{equation}

We now turn to the MMW relation at next-to-leading order in the coupling constant, where we expect the constant on the right-hand side of (\ref{eq:MMW3}) to vary with the coupling strength. Here we will follow \cite{Ghiglieri:2018dib} and show the results as function of the ratio $\mD/T$, where the Debye mass is given by
\begin{equation}
\mD^2 = \left( 1 + \frac{N_f}{6} \right) g^2T^2 \, .
\label{eq:Debye}
\end{equation}
The perturbative expansion of the Debye screening mass $\mD^2$ encounters a nonperturbative contribution at order $g^3T^2$ \cite{Rebhan:1993az,Arnold:1995bh}. A rigorous definition of the Debye mass can be obtained in the framework of effective field theory and dimensional regularization \cite{Braaten:1995jr}. The NLO correction in this framework appears at order $g^4T^2$ and is small over the entire range of couplings considered here ($\mD/T<2.5$ corresponding to $0<\alpha_s<1/\pi$). Since the NLO contributions to $\eta/s$ and $\qT$ appear at order $g$ relative to the lowest-order results, it is consistent to ignore the NLO contribution to the Debye mass.

The jet quenching parameter at NLO is given by \cite{CaronHuot:2008ni,Ghiglieri:2015ala}:
\begin{equation}
\hat{q}_R^\mathrm{(NLO)} = \hat{q}_R^\mathrm{(LO)} +  \delta\hat{q}_R 
\label{eq:qhat}
\end{equation}
with $\hat{q}_R^\mathrm{(LO)}$ given by \cite{Arnold:2008vd}:
\begin{multline}
\hat{q}_R^\mathrm{(LO)} = \frac{g^4 C_R T^3}{2\pi^3} \left[ \left(6+\frac{3}{2}N_f\right) \zeta(3) \ln\frac{\qmax}{\mD} \right.
\\
+ (6+N_f)~(\zeta(2)-\zeta(3)) \times
\\
\times\left( \theta(T-\mD)~\ln\frac{T}{\mD} +\frac{1}{2} - \gamma_\mathrm{E} +\ln 2\right)
 \\
\left. \phantom{\frac{1}{2}} - 6\sigma_{+} - 2N_f\sigma_{-}  \right]
\label{eq:qhatLO}
\end{multline}
and
\begin{equation}
\delta\hat{q}_R = \frac{g^4 C_R C_A T^2 \mD}{32 \pi^2} (3\pi^2 + 10 - 4\ln 2) \, .
\label{eq:dqhatNLO}
\end{equation}
Here $C_A=N_c$ is the quadratic Casimir for the adjoint representation. 

Several comments are in order concerning the lowest-order (LO) result (\ref{eq:qhatLO}). First, the leading logarithm contains an ultraviolet cut-off $\qmax$ analogous to the cut-off in the LL result (\ref{eq:qhatLL}). For a given process involving the transverse momentum diffusion of an energetic parton, e.\ g.\ jet quenching, it should be chosen at the boundary between multiple scattering and isolated hard scatterings in the medium. Here we will continue to to set $\qmax = \mu$; we will clarify this choice explicitly by the notation $\hat{q}(\mu)$. When $\hat{q}$ is deduced from experimental data by model-data comparison, the empirical value will need to be corrected for the specific choice of the momentum transition between transverse diffusion and Coulomb scattering made in the model calculation. 

A second comment concerns the term multiplied by a step function. This term accounts for the difference in the effective final-state density of medium particles after soft scattering ($q_\perp < T$) and hard thermal scattering ($T < q_\perp < \qmax$). In the sub-thermal momentum transfer domain final-state quantum effects (Bose enhancement or Pauli blocking) must be taken into account; for epithermal scatterings these quantum effects are negligible. When $\mD > T$, which occurs for $g > 1$, the low-momentum domain does not contribute a logarithmic term. Although the power counting is questionable in this range of couplings, we bravely extrapolate into this domain here in order to make contact with real-world quark-gluon plasma conditions.

\begin{figure}[htb]
	\centering
	\includegraphics[width=0.95\linewidth]{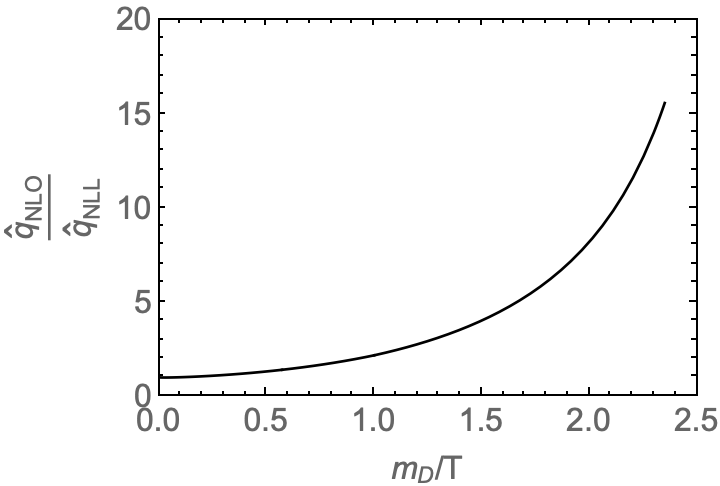}
	\caption{The ratio of $\hat{q}^\mathrm{(NLO)}/\hat{q}^\mathrm{(LL)}$ as function of $\mD/T$.}
	\label{fig:qhat-ratio}
\end{figure}

Figure \ref{fig:qhat-ratio} shows the ratio $\hat{q}^\mathrm{(NLO)}(\mu)/\hat{q}^\mathrm{(LL)}(\mu)$ as a function of $\mD/T$. The NLO contribution is large and leads to an increase by a factor 15 at the upper end of the considered coupling constant range.

The NLO result for the shear viscosity was derived in \cite{Ghiglieri:2018dib} within the effective kinetic theory framework for thermal gauge theories developed by Arnold, Moore and Yaffe \cite{Arnold:2002zm} and its NLO extension by Ghiglieri, Moore and Teaney \cite{Ghiglieri:2015ala}, which was also used for the NLO calculation of $\hat{q}$. The NLO corrections to $\eta$ occur at $O(g)$ and derive from diagrams that contain infrared enhancements from soft, $O(gT)$, thermal interactions. Gluon-mediated contributions have a physical interpretation as diffusion processes; quark-mediated enhancements correspond to conversion processes in which a thermal parton changes its flavor quantum number. The soft interactions can be expressed in terms of light-front correlators, $\hat{q}$ being an example, which can be resummed using Fokker-Planck equations. 

The NLO calculation for $\eta/s$ incorporates NLO corrections to the transverse and longitudinal momentum diffusion coefficients, $\hat{q}$ and $\hat{q}_L$, the quark-gluon conversion rate, and the collinear $1 \leftrightarrow 2$ splitting processes. In addition, one must carefully identify and correct for regions in which the NLO diffusion coefficients become negative or the assumption of strictly collinear splitting breaks down. We refer to \cite{Ghiglieri:2018dib} for the intricate technical details involved the calculation.

The NLO result for $\eta$ does not have a simple analytical representation. However, Ghiglieri, {\em et al.} \cite{Ghiglieri:2018dib} provided an analytical fit to their result, which we are using here:
\begin{multline}
\eta_\mathrm{NLO}^\mathrm{fit} = \left[ \frac{g^4}{\eta_1 T^3} \left( \frac{1}{b} \ln\left( a+(\mu_*/\mD)^b \right) \right. \right.
\\
\left. \left. + \frac{d}{(1+\mD/T)^3} \right) + \frac{\mD}{\eta^{\delta\hat{q}}(cT+\mD)} \right]^{-1}
\label{eq:etaNLO}
\end{multline}
with numerical coefficients $\eta_1, \mu_*/T, a, b, c, d$ given in Table~\ref{tab1} for $N_f=0$ and $N_f=3$ and
\begin{equation}
\eta^{\delta\hat{q}} = \frac{(2\pi)^4T^6}{945} \left( \frac{2(N_c^2-1)}{\delta\hat{q}_A}+\frac{31\,N_f N_c}{8\,\delta\hat{q}_F} \right) \, ,
\label{eq:etadq}
\end{equation}
where $\delta\hat{q}_R$ is given by (\ref{eq:dqhatNLO}).

\begin{table}[htb]
\centering
\begin{tabular}{|c|c|c|}
\hline
   & $N_f=0$ & $N_f=3$ \cr
\hline
   $\eta_1$ & 27.126 & 106.664 \cr
   $\mu_*/T$ & 2.765 & 2.957 \cr
   $a$ & 8.5176 & 4.45096 \cr
   $b$ & 1.38936 & 1.2732 \cr
   $c$ & 1.66144 & 1.91568 \cr
   $d$ & -0.100421 & -0.0777985 \cr
\hline
\end{tabular}
\caption{Numerical constants in the analytical fit (\ref{eq:etaNLO}) to the NLO shear viscosity for a pure gluon plasma ($N_f=0$) and a three-flavor quark-gluon plasma ($N_f=3$) \cite{Ghiglieri:2018dib}.}
\label{tab1}
\end{table}

\begin{figure}[htb]
	\centering
	\includegraphics[width=0.95\linewidth]{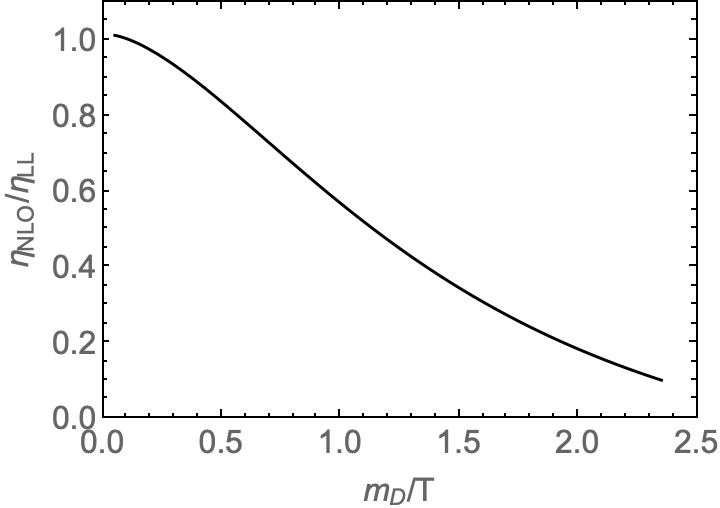}
	\caption{The ratio of $\eta_\mathrm{NLO}/\eta_\mathrm{LL}$ for a three-flavor quark-gluon plasma as function of $\mD/T$.}
	\label{fig:eta-ratio}
\end{figure}

Figure \ref{fig:eta-ratio} shows the ratio $\eta_\mathrm{NLO} / \eta_\mathrm{LL}$ for a three-flavor ($N_f=3$) quark-gluon plasma as a function of $\mD/T$. The NLO contribution is again large, but here it leads to a strong suppression by more than a factor 10 at the upper end of the considered coupling constant range. 

\begin{figure}[htb]
	\centering
	\includegraphics[width=0.95\linewidth]{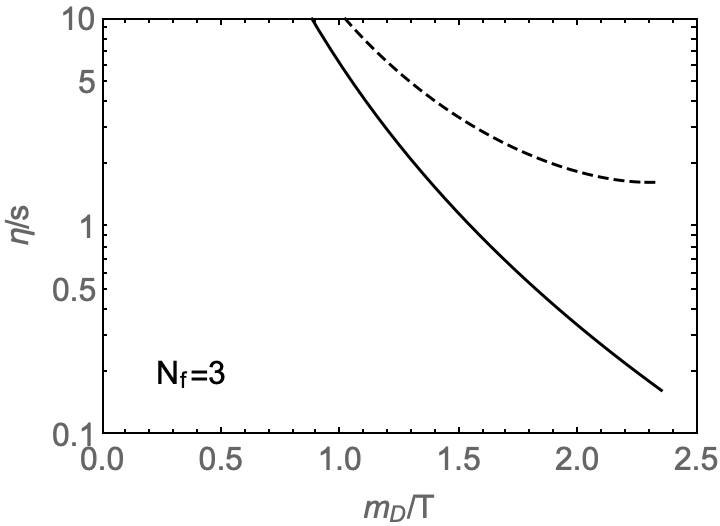}
	\caption{The specific shear viscosity $\eta/s$ at next-to-leading order (solid line) and leading-logarithmic order (dashed line) 
	for a three-flavor quark-gluon plasma as function of $\mD/T$.}
	\label{fig:etas}
\end{figure}

NLO corrections to the entropy density $s$ appear only at order $g^2$. For consistency we therefore follow Ghiglieri, {\em et al.} \cite{Ghiglieri:2018dib} and do not consider those correction when calculating $\eta/s$ at next-to-leading order. In Fig.~\ref{fig:etas} we show $(\eta/s)_\mathrm{NLO}$ (solid line) and $(\eta/s)_\mathrm{LL}$ (dashed line) separately as function of $\mD/T$. At the upper end of the range of coupling constants considered here, the NLO result is comparable to the value deduced from experimental data \cite{Bernhard:2019bmu,Everett:2020xug}.

The large difference between the LL results and NLO results for both transport coefficients naturally raises the question of the range of coupling constants over which the hard-thermal loop (HTL) improved perturbative series converges. For values of the coupling constant realized in the quark-gluon plasma ($\alpha_s \approx 0.3$) the NLO results differ from the LL  results by more than a factor 10. The NNLO correction might result in an even larger change. While only an explicit calculation can answer this question with certainty, there are reasons to believe that the large relative size of the NLO correction is an artifact of the exceptionally small (large) leading order result for $\hat{q}$ ($\eta$). 

\begin{figure}[htb]
	\centering
	\includegraphics[width=0.95\linewidth]{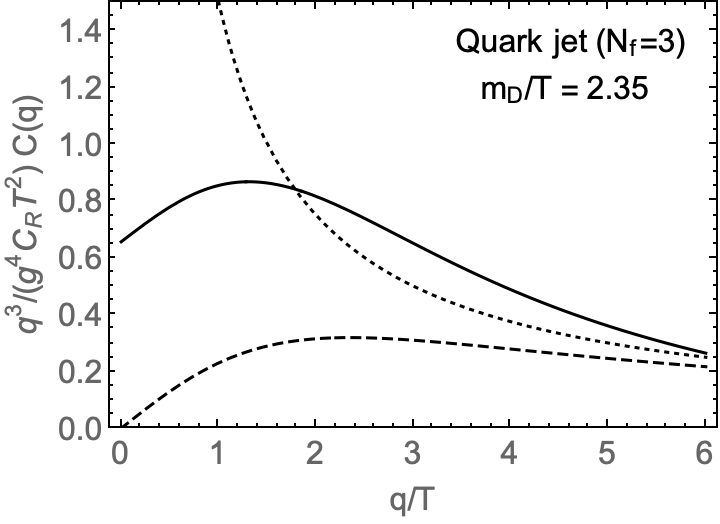}
	\caption{The momentum transfer-weighted collision kernel $q_\perp^3/(g^4C_R T^2) C(q_\perp)$ as a function of $q_\perp/T$ 
	for a quark jet in a three-flavor quark-gluon plasma for $\mD/T = 2.35$. The lowest-order result is shown as a dashed line; 
	the solid curve shows the NLO result. The dotted curve indicates the unscreened collision kernel.}
	\label{fig:q3Cq}
\end{figure}

In order to understand the reason for this, it is useful to consider the difference between the LO and NLO expressions for the two-body collision kernel $C(q_\perp)$ (see Fig~1 in \cite{CaronHuot:2008ni}). Figure \ref{fig:q3Cq} shows the momentum transfer-weighted collision kernel $q_\perp^3/(g^4C_R T^2) C(q_\perp)$ as a function of $q_\perp/T$ for $\mD/T = 2.35$ ($\alpha_s \approx 0.28$). The LO result is shown as a dashed line; the solid curve shows the NLO result. The large increase in the value of $\hat{q}$ arises at small values of $q_\perp/T$. While $q_\perp^3 C(q_\perp)$ tends to zero for $q_\perp \to 0$ in lowest order, it assumes a sizable finite value at next-to-leading order. This indicates that the infrared suppression of elastic two-body scattering is overestimated in the screening corrected LO calculation. 

To wit, the lowest-order result for $q_\perp^3 C(q_\perp)$ in the absence of thermal screening, shown by the dotted curve in Fig.~\ref{fig:q3Cq}, diverges as $1/q_\perp$ at small $q_\perp$, while the screened LO result tends to zero. Seen from this perspective, the NLO result, which incorporates the full dynamical screening of the one-gluon exchange process (diagram (b) in Fig.~3 of \cite{CaronHuot:2008ni}), just corrects for the unphysically large suppression in the infrared limit caused by static screening. This physical effect first appears at NLO and leads to a large change in the value of the transport coefficients. Put differently, the LL result for $\hat{q}$ is unphysically small, because the suppression of the collision kernel near $q_\perp \to 0$ is quadratic in $q_\perp$ when it should only be linear in $q_\perp$. Whether this argument is, indeed, correct can only be answered by a NNLO calculation of the collision kernel, which is cumbersome because of the proliferation of Feynman diagrams but does not appear to be impractical.

\begin{figure}[htb]
	\centering
	\includegraphics[width=0.95\linewidth]{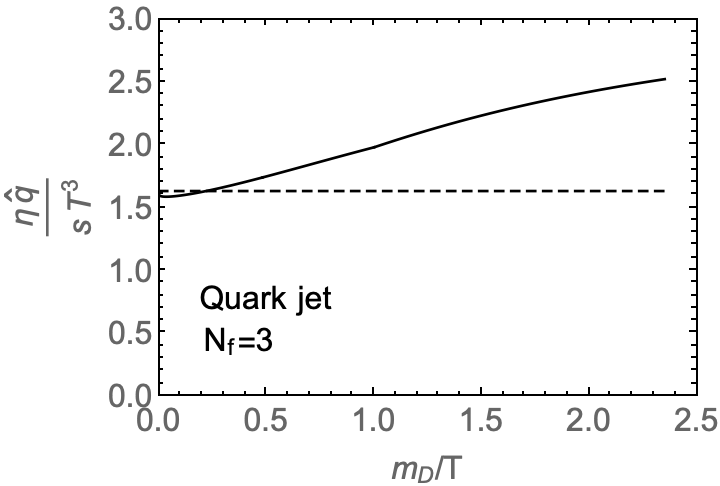}
	\caption{The double ratio $(\eta/s)(\hat{q}(\mu)/T^3)$ for a three-flavor quark-gluon plasma and a quark jet as function of $\mD/T$. 
	The NLO result is shown as the solid line; the leading-logarithmic result (\ref{eq:MMW3}) is shown by the dashed line.}
	\label{fig:MMW-3}
\end{figure}

\begin{figure}[htb]
	\centering
	\includegraphics[width=0.95\linewidth]{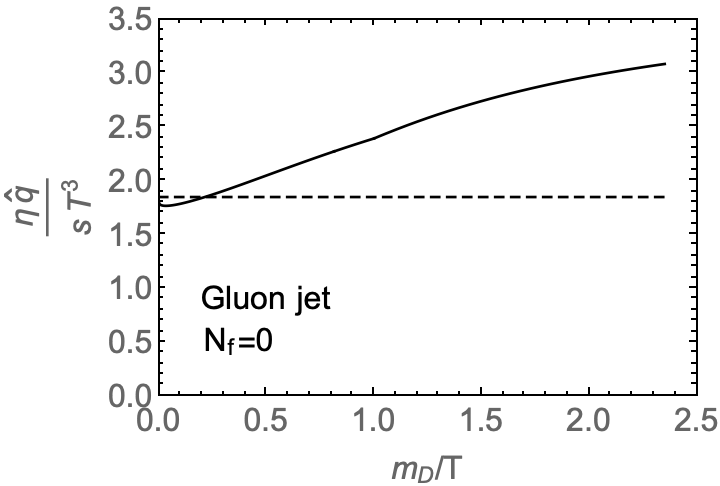}
	\caption{The double ratio $(\eta/s)(\hat{q}(\mu)/T^3)$ for a pure gluon plasma and a gluon jet as function of  $\mD/T$. 
	The NLO result is shown as the solid line; the leading-logarithmic result (\ref{eq:MMWg}) is shown by the dashed line.}
	\label{fig:MMW-g}
\end{figure}

Finally, we are ready to evaluate the double ratio $(\eta/s)(\hat{q}(\mu)/T^3)$ at next-to-leading order. The results are shown as solid lines in Fig.~\ref{fig:MMW-3} for a three-flavor quark-gluon plasma ($N_f=3$) and a quark-jet, and in Fig.~\ref{fig:MMW-g} for a pure gluon plasma and a gluon jet. The dashed lines in both figures show the coupling constant-independent result at LL order. The figures show that the large NLO corrections to the two dimensionless ratios, $\eta/s$ and $\hat{q}/T^3$, cancel to a large extent. At realistic couplings ($\mD/T \sim 2-2.5$) the NLO result for the double ratio differs from the LL result only by a factor $\sim 1.6$ and is weakly dependent on the coupling strength.

The results shown in Figs.~\ref{fig:MMW-3} and \ref{fig:MMW-g} confirm the expectation that the MMW relation (\ref{eq:MMW}) is rather robust against next-to-leading order corrections to the transport coefficients $\eta$ and $\hat{q}$. This expectation was based on the generic argument that such a relation arises in any theory in which transport processes are carried by quasiparticles and soft interactions dominate the total scattering cross section. 

\begin{figure}[htb]
	\centering
	\includegraphics[width=0.95\linewidth]{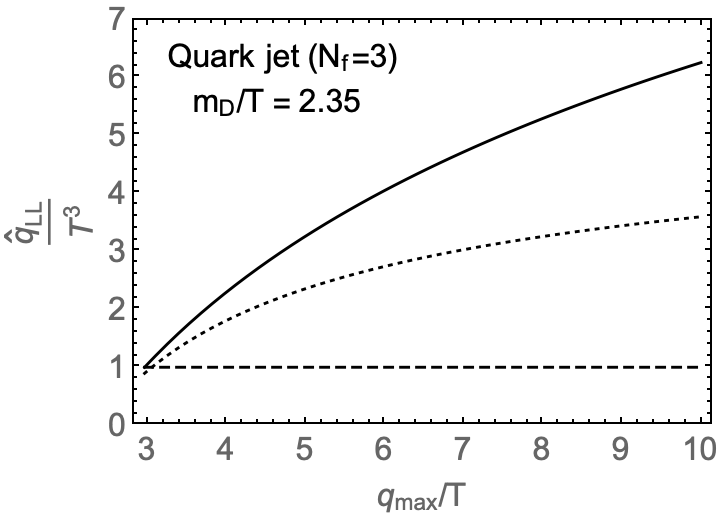}
	\\
	\includegraphics[width=0.95\linewidth]{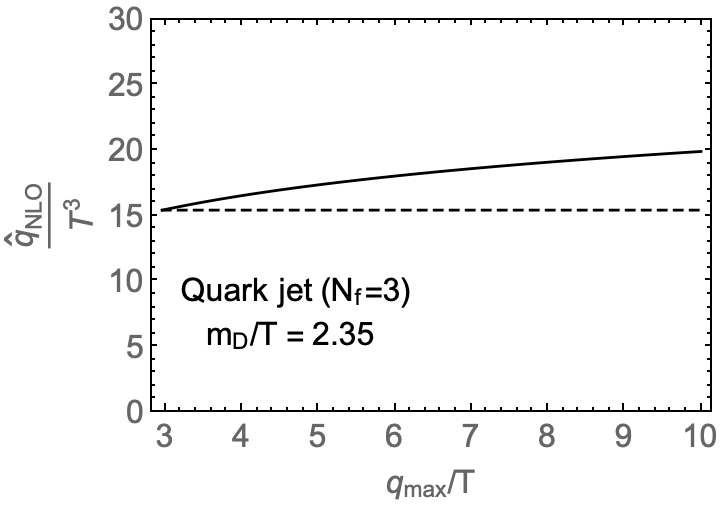}
	\caption{Upper panel: Dependence of $\hat{q}^\mathrm{(LL)}(\qmax)/T^3$ at leading-logarithmic order on the 
	high-momentum cut-off $\qmax$ as function of $\qmax/T$ (solid line) together with the result 
	(\ref{eq:qhatLL}) for $\qmax/T = \mu/T = 2.975$ (dashed line). The dotted line shows the cut-off dependence 
	when the running of the coupling constant is taken into account (see (2.8) in \cite{Arnold:2008vd}). 
	Lower panel: Dependence of $\hat{q}^\mathrm{(NLO)}(\qmax)/T^3$ at next-to-leading order on the 
	high-momentum cut-off $\qmax$ as function of $\qmax/T$ (solid line) together with the NLO result 
	(\ref{eq:qhat}) for $\qmax/T = \mu/T = 2.975$ (dashed line).}
	\label{fig:qhatT3}
\end{figure}

One issue we have so far largely ignored is that the value of $\hat{q}$ depends on the ultraviolet cut-off $\qmax$. The upper panel of Fig.~\ref{fig:qhatT3} shows the cut-off dependence of the leading-log result $\hat{q}^\mathrm{(LL)}(\qmax)/T^3$ as function of $\qmax/T$ (solid line) together with the result (\ref{eq:qhatLL}) for $\qmax = \mu$ (dashed line). The figure shows that the value for $\hat{q}$ depends strongly on the chosen value of the momentum cut-off $\qmax$. The dotted line shows the cut-off dependence of $\hat{q}^\mathrm{(LL)}(\qmax)/T^3$ when the running of the coupling constant is taken into account (see eq.~(2.8) in \cite{Arnold:2008vd}). Since the NLO correction is large and independent of a high-momentum cut-off, the relative magnitude of the cut-off dependence at next-to-leading order is much weaker, as shown in the lower panel of Fig.~\ref{fig:qhatT3}, even without the running coupling modification. This demonstrates that the NLO result for the double ratio $(\eta/s)(\hat{q}/T^3)$ is much more robust against the choice of the high-momentum cut-off $\qmax$ than the lowest-order (LL or LO) result.

In conclusion, the double ratio of quark-gluon plasma transport coefficients, $(\eta/s)(\hat{q}(\mu)/T^3)$, has been shown to be robust against higher-order corrections to the individual transport coefficients, which are large. At next-to-leading order, the double ratio is much less sensitive to the high-momentum cut-off of the jet quenching parameter $\hat{q}$ than at lowest order. Our result provides compelling motivation for a {\em simultaneous} extraction of both, $\eta/s$ and $\hat{q}/T^3$, from experimental data by a state-of-the-art model-data comparison.

{\em Acknowledgments:} I thank J.-F.~Paquet for multiple helpful discussions and A.~Majumber for valuable comments on a draft of the manuscript. This work was supported by the Office of Science of the U.~S.~Department of Energy under Grant DE-FG02-05ER41367.


\begin{thebibliography}{99}

\bibitem{Bernhard:2019bmu}
J.~E.~Bernhard, J.~S.~Moreland and S.~A.~Bass,
%``Bayesian estimation of the specific shear and bulk viscosity of quark\textendash{}gluon plasma,''
Nature Phys. \textbf{15}, no.11, 1113-1117 (2019).
%doi:10.1038/s41567-019-0611-8

\bibitem{Everett:2020xug}
D.~Everett \textit{et al.} [JETSCAPE],
%``Multisystem Bayesian constraints on the transport coefficients of QCD matter,''
Phys. Rev. C \textbf{103}, no.5, 054904 (2021)
%doi:10.1103/PhysRevC.103.054904
[arXiv:2011.01430 [hep-ph]].

\bibitem{Burke:2013yra}
K.~M.~Burke \textit{et al.} [JET Collaboration],
%``Extracting the jet transport coefficient from jet quenching in high-energy heavy-ion collisions,''
Phys. Rev. C \textbf{90}, no.1, 014909 (2014)
%doi:10.1103/PhysRevC.90.014909
[arXiv:1312.5003 [nucl-th]].

\bibitem{Cao:2021keo}
S.~Cao, \textit{et al.} [JETSCAPE Collaboration],
%``Determining the jet transport coefficient $\hat{q}$ from inclusive hadron suppression measurements using Bayesian parameter estimation,''
[arXiv: 2102.11337 [nucl-th]].

\bibitem{Majumder:2007zh}
A.~Majumder, B.~M\"uller and X.~N.~Wang,
%``Small shear viscosity of a quark-gluon plasma implies strong jet quenching,''
Phys. Rev. Lett. \textbf{99}, 192301 (2007)
%doi:10.1103/PhysRevLett.99.192301
[arXiv:hep-ph/0703082 [hep-ph]].

%\cite{CaronHuot:2008ni}
\bibitem{CaronHuot:2008ni}
S.~Caron-Huot,
%``O(g) plasma effects in jet quenching,''
Phys. Rev. D \textbf{79}, 065039 (2009)
%doi:10.1103/PhysRevD.79.065039
[arXiv:0811.1603 [hep-ph]].

\bibitem{Ghiglieri:2015ala}
J.~Ghiglieri, G.~D.~Moore and D.~Teaney,
%``Jet-Medium Interactions at NLO in a Weakly-Coupled Quark-Gluon Plasma,''
JHEP \textbf{03}, 095 (2016)
%doi:10.1007/JHEP03(2016)095
[arXiv:1509.07773 [hep-ph]].

\bibitem{Ghiglieri:2018dib}
J.~Ghiglieri, G.~D.~Moore and D.~Teaney,
%``QCD Shear Viscosity at (almost) NLO,''
JHEP \textbf{03}, 179 (2018)
doi:10.1007/JHEP03(2018)179
[arXiv:1802.09535 [hep-ph]].

\bibitem{Baier:1996kr}
R.~Baier, Y.~L.~Dokshitzer, A.~H.~Mueller, S.~Peigne and D.~Schiff,
%``Radiative energy loss of high-energy quarks and gluons in a finite volume quark - gluon plasma,''
Nucl. Phys. B \textbf{483}, 291-320 (1997)
%doi:10.1016/S0550-3213(96)00553-6
[arXiv:hep-ph/9607355 [hep-ph]].

\bibitem{Danielewicz:1984ww}
P.~Danielewicz and M.~Gyulassy,
%``Dissipative Phenomena in Quark Gluon Plasmas,''
Phys. Rev. D \textbf{31}, 53-62 (1985).
%doi:10.1103/PhysRevD.31.53

\bibitem{Arnold:2000dr}
P.~B.~Arnold, G.~D.~Moore and L.~G.~Yaffe,
%``Transport coefficients in high temperature gauge theories. 1. Leading log results,''
JHEP \textbf{11}, 001 (2000)
%doi:10.1088/1126-6708/2000/11/001
[arXiv:hep-ph/0010177 [hep-ph]].

\bibitem{Rebhan:1993az}
A.~K.~Rebhan,
%``The NonAbelian Debye mass at next-to-leading order,''
Phys. Rev. D \textbf{48}, R3967-R3970 (1993)
%doi:10.1103/PhysRevD.48.R3967
[arXiv:hep-ph/9308232 [hep-ph]].

\bibitem{Arnold:1995bh}
P.~B.~Arnold and L.~G.~Yaffe,
%``The NonAbelian Debye screening length beyond leading order,''
Phys. Rev. D \textbf{52}, 7208-7219 (1995)
%doi:10.1103/PhysRevD.52.7208
[arXiv:hep-ph/9508280 [hep-ph]].

\bibitem{Braaten:1995jr}
E.~Braaten and A.~Nieto,
%``Free energy of QCD at high temperature,''
Phys. Rev. D \textbf{53}, 3421-3437 (1996)
%doi:10.1103/PhysRevD.53.3421
[arXiv:hep-ph/9510408 [hep-ph]].

\bibitem{Arnold:2008vd}
P.~B.~Arnold and W.~Xiao,
%``High-energy jet quenching in weakly-coupled quark-gluon plasmas,''
Phys. Rev. D \textbf{78}, 125008 (2008)
%doi:10.1103/PhysRevD.78.125008
[arXiv:0810.1026 [hep-ph]].

\bibitem{Arnold:2002zm}
P.~B.~Arnold, G.~D.~Moore and L.~G.~Yaffe,
%``Effective kinetic theory for high temperature gauge theories,''
JHEP \textbf{01}, 030 (2003)
%doi:10.1088/1126-6708/2003/01/030
[arXiv:hep-ph/0209353 [hep-ph]].

\end{thebibliography}
\end{document}